\documentclass[runningheads]{llncs}
\usepackage{refs}
\usepackage[frozencache,cachedir=.]{minted}
\usepackage{comment}

\usepackage{amsmath}
\usepackage{minted}
\usepackage{listings}
\usepackage{graphicx}
\usepackage{xcolor}
\usepackage{subcaption}
\usepackage{wrapfig}

\usepackage{amsfonts}

\usepackage{multirow}
\usepackage[normalem]{ulem}

\newcommand{\Op}{{\textit{Op}}}

\newcommand{\Omit}[1]{}
\newcommand{\twr}[1]{{\color{magenta}\emph{T: #1}}}

\newcommand{\minitab}[2][l]{\begin{tabular}{#1}#2\end{tabular}}

\usepackage{graphicx}                   
\usepackage{hyperref}                   
\usepackage[firstpage]{draftwatermark}  

\usepackage{xspace}

\newcommand{\sys}{\textsc{Quasimodo}\xspace}

\begin{document}
\title{Symbolic Quantum Simulation with \sys}
\author{Meghana Sistla\inst{1} \and 
Swarat Chaudhuri\inst{1} \and
Thomas Reps\inst{2} }
\institute{The University of Texas at Austin \\
\email{mesistla@utexas.edu, swarat@cs.utexas.edu} 
\and
University of Wisconsin-Madison \\
\email{reps@cs.wisc.edu}
}
\date{January 2023}
\maketitle

\SetWatermarkAngle{0}
\SetWatermarkText{\raisebox{14.5cm}{%
  \hspace{0.15cm}%
  \href{https://doi.org/10.5281/zenodo.7922448}{\includegraphics{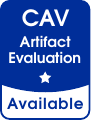}}%
  \hspace{9cm}%
  \includegraphics{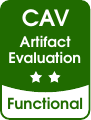}%
 }
}

\begin{abstract}
    The simulation of quantum circuits on classical computers is an important problem in quantum computing. Such simulation requires representations of distributions over very large sets of basis vectors, and recent work has used symbolic data-structures such as Binary Decision Diagrams (BDDs) for this purpose. 
    In this tool paper, we present \sys, an extensible, open-source Python library for \emph{symbolic simulation} of quantum circuits. 
    \sys is specifically designed for easy extensibility to other backends.
    \sys allows simulations of quantum circuits, checking properties of the outputs of quantum circuits, and debugging quantum circuits.     
    It also allows the user to choose from among several symbolic data-structures---both unweighted and weighted BDDs, and a recent structure called 
    Context-Free-Language Ordered Binary Decision Diagrams 
    (CFLOBDDs)---and can be easily extended to support other symbolic data-structures.
    
\end{abstract}
\section{Introduction}
\label{Se:Introduction}

Canonical, symbolic representations of Boolean functions---for example, Binary Decision Diagrams (BDDs) \cite{toc:Bryant86}---have a long history in automated system design and verification.
More recently, such data-structures have found exciting new applications in \emph{quantum simulation}.
Quantum computers can theoretically solve certain problems much faster than traditional computers, but current quantum computers are error-prone and
access to them is limited.
The simulation of quantum algorithms on classical machines allows researchers to experiment with quantum algorithms even without
access to
reliable hardware.

Symbolic function representations are helpful in quantum simulation because a quantum system's state can be viewed as a distribution over an exponential-sized set of basis-vectors (each representing a ``classical'' state).
Such a state, as well as transformations that quantum algorithms typically apply to them, can
often
be efficiently represented using a symbolic data-structure.
Simulating an algorithm then amounts to performing a sequence of symbolic operations. 


Currently, there are a small number of open-source software systems that support such \emph{symbolic quantum simulation} \cite{willeVisualizingDecisionDiagrams2021,Qiskit,cirq_developers_2022_7465577,gray2018quimb,9586191}.
However, the underlying symbolic data-structure can have an enormous effect on simulation performance.
In this tool paper, we present \sys,\footnote{
  \sys is available at 
\url{https://github.com/trishullab/Quasimodo.git}.
}
an extensible framework for symbolic quantum simulation.
\sys is specifically designed for easy extensibility to other backends to make it possible to experiment with a variety of symbolic data-structures.
\sys currently supports (i) BDDs \cite{toc:Bryant86,DBLP:journals/fmsd/FujitaMY97,DBLP:journals/fmsd/BaharFGHMPS97}, (ii) a weighted variant of BDDs \cite{DBLP:journals/tcad/NiemannWMTD16,DBLP:conf/date/ViamontesMH04}, \cite[Ch.\ 5]{Book:ZW2020}, and 
(iii) Context-Free-Language Ordered Binary Decision Diagrams
CFLOBDDs \cite{sistla2022cflobdds}, a recent canonical representation of Boolean functions that has been shown to outperform BDDs in many quantum-simulation tasks.
\sys also has a clean interface that formal-methods researchers can use to plug in new symbolic data-structures,
which helps to lower the barrier to entry for formal-methods researchers interested in this area.


Users access \sys through a Python interface.
They can define a quantum algorithm as a quantum circuit using 18 different kinds of quantum gates, such as Hadamard, CNOT, and Toffoli gates.
They can simulate the algorithm using a symbolic data-structure of their own choosing.
Users can sample outcomes from the probability distribution computed through simulation, and can query the simulator for the probability of a specific outcome of a quantum computation over a set of quantum bits (qubits).
The system also allows for a form of correctness checking: users are allowed to ask for the set of \emph{all} high-probability outcomes and to check that these satisfy a given assertion. 


Along with \sys, we are releasing a suite of 7 established quantum algorithms encoded in the input language of \sys.
We hope that these algorithms will serve as benchmarks for future research on symbolic simulation and verification of quantum algorithms.


\smallskip
\noindent
\textit{Organization.}
\S\ref{Se:quantumsim} gives an overview of quantum simulation.
\S\ref{Se:tool} gives a user-level overview of \sys.
\S\ref{Se:Background} provides background on the symbolic data-structures available in \sys.
\S\ref{Se:experiments} describes the programming model of \sys, and presents experimental results.
\S\ref{Se:conclusion} concludes.


\section{Background on Quantum Simulation}
\label{Se:quantumsim}


Quantum algorithms on quantum computers can achieve polynomial to exponential speed-ups over classical algorithms on specific problems.
However, because so far there are no practical scalable quantum computers, simulation of quantum circuits on classical computers can help in understanding how quantum algorithms work and scale.
A simulation of a quantum-circuit computation \cite{sistla2022cflobdds,Book:ZW2020,9586191,gray2018quimb,cirq_developers_2022_7465577,Qiskit} uses a representation $qs$ of a quantum state and performs operations on $qs$ that correspond to quantum-circuit operations (gate applications and measurements on $qs$).

Simulating a quantum circuit can have advantages compared to executing the circuit on a quantum computer.
For instance, some quantum algorithms perform multiple iterations of a particular quantum operator $\Op$ (e.g., $k$ iterations, where $k = 2^j$).
A simulation can operate on $\Op$ itself \cite[Ch.\ 6]{Book:ZW2020}, using $j$ iterations of repeated squaring to create matrices for $\Op^2$, $\Op^4, \ldots$, $\Op^{2^j} = \Op^k$.
In contrast, a physical device must apply $\Op$ sequentially, and thus performs $\Op$ $k = 2^j$ times.

Many quantum algorithms require multiple measurements on the final state.
After a measurement on a quantum computer, the quantum state collapses to the measured state.
Thus, every successive measurement requires re-running the quantum circuit.
However, with a simulation, the quantum state can be preserved across measurements, and thus the quantum circuit need only be executed once.

\section{\sys's Programming and Analysis Interface}
\label{Se:tool}

\begin{figure}[tb!]
    \begin{minted}[linenos, escapeinside=!!]{python}
        import quasimodo #python package to import for Quasimodo !\label{Li:importPackage}!
        epsilon = 1e-8

        # number of qubits in the quantum state
        numQubits = 2 ** 12
        # initialize the quantum state
        qs = quasimodo.QuantumState("CFLOBDD", numQubits)  !\label{Li:QuantumState}! 
        qs.h(0) # Apply Hadamard gate to Qubit 0  !\label{Li:HadamardGate}!
        for i in range(1, numQubits):
            qs.cx(0, i) # Apply CNOT Gate from Qubit 0 to Qubit i  !\label{Li:CnotGate}!

        qubit_mapping = {} # map from qubit number -> desired outcome
        for i in range(0, numQubits):
            qubit_mapping[i] = 1
        
        # query probability of outcome as encoded in qubit mapping
        prob = qs.prob(qubit_mapping) !\label{Li:queryMeasurementProbability}!
        if (abs(prob - 0.5)) < epsilon:
            print ("Circuit is correct")
        else
            print ("Incorrect circuit")
    \end{minted}
    \caption{
    An example of a \sys program that performs a quantum-circuit computation in which the final quantum state is a GHZ state with 4,096 qubits.
    The program verifies that a measurement of the final quantum state has a 50\% chance of returning the all-ones
    basis-state.
    }
    \label{Fi:QuantumExample}
\end{figure}

\noindent
This section presents an overview of \sys from the perspective of a user of the Python API.
A user can define a quantum-circuit computation and check the properties of the quantum state at various points in the computation.
This section also explains how \sys can be easily extended to include custom representations of the quantum state.


\paragraph{\textbf{Example.}} ~\figref{QuantumExample} shows an example of a quantum-circuit computation written using the \sys API.
To use the \sys library, one needs to import the package, as shown in~\linenoref{importPackage}.
A user can then create a program that implements a quantum-circuit computation by
\begin{itemize}
  \item
    Initializing the quantum state by making a call to ${\tt QuantumState}$ with an argument that selects the desired backend data-structure and the number of qubits in the quantum state.
    (See \linenoref{QuantumState}.)
    The example in 
    \figref{QuantumExample} uses CFLOBDD as the backend simulator, but other data-structures can be used by changing the backend parameter to BDD or WBDD.
    ${\tt QuantumState}$ sets the initial quantum state to the all-zeros basis-state.
  \item
    Applying single-qubit gates to the quantum state, such as Hadamard (h), Pauli-X (x), T-Gate (t), and others.
    The qubit to which they are to be applied is specified by passing the qubit number.
    (See \linenoref{HadamardGate}.)
  \item
    Applying multi-qubit gates to the quantum state, such as CNOT (cx), Toffoli (ccx), SWAP (swap), and others.
    The qubits to which they are to be applied is specified by passing the qubit numbers.
    (See \linenoref{CnotGate}.)
\end{itemize}

Note that queries on the quantum state do not have to be made only at the end of the program;
they can also be interspersed throughout the circuit-simulation computation.

\sys allows different backend data-structures to be used for representing quantum states.
It comes with BDDs \cite{toc:Bryant86,DBLP:journals/fmsd/FujitaMY97,DBLP:journals/fmsd/BaharFGHMPS97}, a weighted variant of BDDs \cite{DBLP:journals/tcad/NiemannWMTD16,DBLP:conf/date/ViamontesMH04}, \cite[Ch.\ 5]{Book:ZW2020}, and CFLOBDDs \cite{sistla2022cflobdds}.
\sys also provides an interface for new backend data-structures to be incorporated by users.
All three of the standard backends provide compressed representations of quantum states and quantum gates, although---as with all variants of decision diagrams---state representations may blow up as a sequence of gate operations are performed.



\paragraph{Quantum Simulation.}
Quantum simulation problems can be implemented using \sys by defining a quantum-circuit computation, and then invoking the API function \texttt{measure} to sample a basis-vector from the final quantum state.
For instance, suppose that the final quantum state is
$\begin{bmatrix}
0.5 & 0 & 0.5 & 0.5 & 0 & 0 & 0.5 & 0
\end{bmatrix}$. Then
${\tt measure}$ would return a string in the set $\{ 000, 010, 011, 110\}$ with probability 0.25 for each of the four strings.

\paragraph{Verification.}
As shown in \linenoref{queryMeasurementProbability} of \figref{QuantumExample}, \sys provides an API call to inquire about the probability of a
specific outcome.
The function ${\tt prob}$ takes as its argument a mapping from qubits to $\{0, 1\}$, which defines a basis-vector $e$ of interest, and returns the probability that the state would be $e$ if a measurement were carried out at that point.
It can also be used to query the probability of a set of outcomes, using a mapping of just a subset $S$ of the qubits, in which case ${\tt prob}$ returns the sum of all probabilities of obtaining a state that satisfies $S$.
For example, if the quantum state computed by a 3-qubit circuit over $\langle q_0, q_1, q_2 \rangle$ is 
$\begin{bmatrix}
0.5 & 0 & 0.5 & 0.5 & 0 & 0 & 0.5 & 0
\end{bmatrix}$,
the user can query the probability of states satisfying $q_1 = 1 \land q_2 = 0$ by calling ${\tt prob({1 : 1, 2 : 0})}$, which would returns 0.5
(= $Pr(q_0 = 0 \land q_1 = 1 \land q_2 = 0) + Pr(q_0 = 1 \land q_1 = 1 \land q_2 = 0)$ = $(0.5)^2$ + $(0.5)^2$).

Given a relational specification $R(x, y)$ and a quantum circuit $y = Q(x)$, this feature is useful for verifying properties of the form ``$Pr[R(x, Q(x))] > \theta$,'' where $\theta$ is some desired probability threshold for the user's application.

\paragraph{Debugging Quantum Circuits.}
\sys additionally provides a feature to query the number of outcomes for a given probability.
This feature is especially helpful for debugging large quantum circuits---large in-terms of qubit counts---when most outcomes have similar probabilities.

Consider the case of a quantum circuit whose final quantum state is intended to be
$
\frac{1}{\sqrt{6}}
\begin{bmatrix}
    1 & 1 & 1 & 0 & 1 & 1 & 1 & 0
\end{bmatrix}
$.
One can check if the final quantum state is the one intended
by querying the number of outcomes that have probability $\frac{1}{6}$. If the returned value is 6, the user can then check if states $011$ and $111$ have probability 0 by calling ${\tt prob(\{0: 0, 1: 1, 2:1\})}$ and
${\tt prob(\{0:1, 1:1, 2:1\})}$,
respectively.
The API function for querying the number of outcomes that have probability $\texttt{p} \displaystyle \pm \epsilon$ is ${\tt measurement\_counts(p, \epsilon)}$.
One can also query the number of outcomes that have probability $\geq \texttt{p}$ by invoking the function ${\tt tail\_counts(p)}$.

\sys's API provides the methods
${\tt get\_state()}$ and ${\tt most\_frequent()}$ to obtain the quantum state (as a pointer to the underlying data-structure) and the outcome with the highest probability, respectively.

\subsection{Extending \sys}

The currently supported symbolic data-structures for representing quantum states and quantum gates are written in C++ with bindings for Python.
All of the current representations implement an abstract C++ class that exposes 
(i) \texttt{QuantumState}, which returns a state object that represents a quantum state,
(ii) eighteen quantum-gate operations,
(iii) an operation for gate composition,
(iv) an operation for applying a gate---either a primitive gate or the result of gate composition---to a quantum state, and
(v) five query operations.
Users can easily extend \sys to add a replacement backend by providing an operation to create a state object, as well as implementations of the seventeen gate operations and three query operations.
Currently, the easiest path is to implement the custom representation in C++ as an implementation of the abstract C++ class used by \sys's standard backends.



\section{The Internals of \sys}
\label{Se:Background}

In this section, we elaborate on the internals of \sys. Specifically, we briefly summarize the BDD, WBDD, and CFLOBDD data-structures that \sys currently supports, and illustrate how \sys performs symbolic simulation using these data-structures. 
For brevity, we illustrate the way \sys uses these data-structures using the example of the Hadamard gate, a commonly used quantum gate, defined by the matrix $H = \frac{1}{\sqrt{2}} \begin{bmatrix}
        1 & 1\\
        1 & -1
    \end{bmatrix}$.

\begin{figure}[tb!]
    \centering
    \begin{subfigure}[t]{0.3\linewidth}
    \includegraphics[width=0.7\linewidth]{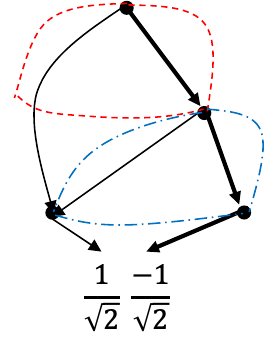}
    \caption{BDD}
    \end{subfigure}
    \vspace{2ex}
    \begin{subfigure}[t]{0.3\linewidth}
    \centering
    \includegraphics[width=\linewidth]{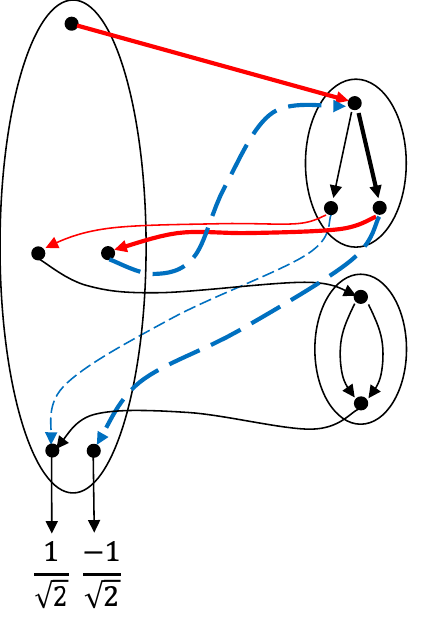}
    \caption{CFLOBDD}
    \end{subfigure}
    \vspace{4ex}
    \begin{subfigure}[t]{0.3\linewidth}
    \centering
    \includegraphics[width=0.7\linewidth]{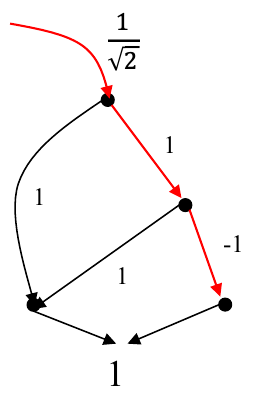}
    \caption{WBDD}
    \end{subfigure}
    \vspace{-0.4in}
    \caption{Three representations of the Hadamard matrix $H = \frac{1}{\sqrt{2}} \begin{bmatrix}
        1 & 1\\
        1 & -1
    \end{bmatrix}$. (a) A BDD, (b) a CFLOBDD, and (c) a WBDD.
    The variable ordering is $\langle x_0, y_0 \rangle$, where $x_0$ is the row decision variable and $y_0$ is the column decision variable.}
    \label{Fi:HRep}
    \vspace{-0.2in}
\end{figure}

\paragraph{Binary Decision Diagrams (BDDs).}
\label{Se:BDDs}

\sys provides an option to use Binary Decision Diagrams (BDDs)~\cite{toc:Bryant86,DBLP:journals/fmsd/FujitaMY97,DBLP:journals/fmsd/BaharFGHMPS97} as the underlying data-structure.
A BDD is a data-structure used to efficiently represent a function from Boolean variables to some space of values (Boolean or non-Boolean).
  The extension of BDDs to support a non-Boolean range is called Multi-Terminal BDDs (MTBDDs) \cite{DBLP:journals/fmsd/FujitaMY97} or Algebraic DDs (ADDs) \cite{DBLP:journals/fmsd/BaharFGHMPS97}.
  In this paper, we use ``BDD'' as a generic term for both BDDs proper and MTBDDs/ADDs.
Each node in a BDD corresponds to a specific Boolean variable, and the node's outgoing edges represents a decision based on the variable's value (0 or 1).
The leaves of the BDD represent the different outputs of the Boolean function.
In the best case, BDDs provide an exponential compression in space compared to the size of the decision-tree representation of the function.\footnote{
  Technically, the BDD variant that, in the best case, is exponentially smaller than the corresponding decision tree, is called a \emph{quasi-reduced BDD}.
  Quasi-reduced BDDs are BDDs in which variable ordering is respected, but don't-care nodes are \emph{not} removed, and thus all paths from the root to a leaf have length $n$, where $n$ is the number of variables.
  However, the size of a quasi-reduced BDD is at most a factor of $n+1$ larger than the size of the corresponding (reduced, ordered) BDD \cite[Thm.\ 3.2.3]{Book:Wegener00}.
  Thus, although BDDs can give better-than-exponential compression compared to decision trees, at best, it is linear compression of exponential compression.
}
\figref{HRep}(a) shows the BDD representation of the Hadamard matrix $H$ with variable ordering $\langle x_0, y_0 \rangle$, where $x_0$ is the row decision variable and $y_0$ is the column decision variable.

We enhanced the CUDD library~\cite{somenzi2012cudd} by incorporating complex numbers at the leaf nodes and adding the ability to count paths.

\paragraph{Context-Free-Language Ordered Binary Decision Diagrams (CFLOBDDs).}
CFLOBDDs~\cite{sistla2022cflobdds} are a binary decision diagram inspired by BDDs, 
but the two data-structures are based on different principles.
A BDD is an acyclic finite-state machine (modulo ply-skipping), whereas a CFLOBDD is a particular kind of \emph{single-entry, multi-exit, non-recursive, hierarchical finite-state machine} (HFSM) \cite{TOPLAS:ABEGRY05}.
Whereas a BDD can be considered to be a special form of bounded-size, branching, but non-looping program, a CFLOBDD can be considered to be a bounded-size, branching, but non-looping program in which a certain form of \emph{procedure call} is permitted.

CFLOBDDs can provide an exponential compression over BDDs and double-exponential compression over the decision-tree representation. The additional compression of CFLOBDDs can be roughly attributed to the following reasons:
\begin{itemize}
    \item As with BDDs, one level of exponential compression comes from sharing in a directed-acyclic-graph
    (i.e., a complete binary tree is folded to a dag).
    \item In CFLOBDDs, there is a further level of exponential compression from reuse of ``procedures'': the same ``procedure'' can be called multiple times at different call sites.
\end{itemize}
Such “procedure calls” allow additional sharing of structure beyond what is possible in BDDs: a BDD can share sub-DAGs, whereas a procedure call in a CFLOBDD shares the ``middle of a DAG''.
The CFLOBDD for Hadamard matrix $H$, shown in~\figref{HRep}(b), illustrates this concept:
the fork node (the node with a split) at the top right of~\figref{HRep}(b) is shared twice---once during the red solid path ({\color{red} ---}) and again during the blue dashed path (${\color{blue}-\cdot-}$).
The corresponding elements of the BDD for $H$ are outlined in red and blue in~\figref{HRep}(a). 
The cell entry $H[1][1]$, which corresponds to the assignment $\{ x_0 \mapsto 1, y_0 \mapsto 1 \}$, is shown in~\figref{HRep}(a) (BDD) and \figref{HRep}(b) (CFLOBDD) as the paths highlighted in bold that lead to the value $\frac{-1}{\sqrt{2}}$.


\paragraph{Weighted Binary Decision Diagrams (WBDDs).}
A Weighted Binary Decision Diagram (WBDD)
\cite{DBLP:journals/tcad/NiemannWMTD16,DBLP:conf/date/ViamontesMH04}, \cite[Ch.\ 5]{Book:ZW2020} is similar to a BDD, but each decision (edge) in the diagram is assigned a weight.
To evaluate the represented function $f$ on a given input $a$ (i.e., $a$ is an assignment in $\{0,1\}^n$), the path for $a$ is followed;
the value of $f(a)$ is the product of the weights encountered along the path.
Consider how the WBDD in \figref{HRep}(c) represents Hadamard matrix $H$.
The variable ordering used is $\langle x_0, y_0 \rangle$, where $x_0$ is the row decision variable and $y_0$ is the column decision variable.
Consider the assignment $a = \{x_0  \mapsto 1, y_0 \mapsto 1\}$.
This assignment corresponds to the path shown in red in~\figref{HRep}(c).
The WBDD has a weight $\frac{1}{\sqrt{2}}$ at the root, which is common to all paths.
The weight corresponding to $\{ x_0 \mapsto 1\}$ is 1 and $\{ y_0 \mapsto 1\}$ is -1;
consequently, $a$ evaluates to $\frac{1}{\sqrt{2}} * 1 * -1 = \frac{-1}{\sqrt{2}}$, which is equal to the value in cell $H[1][1]$.

WBDDs have been used in a variety of applications, such as verification and quantum simulation~\cite{Book:ZW2020}.
In the case of quantum simulation, the weights on the edges of a WBDD are complex numbers\Omit{, and multiplication is used as the composition operator}.
Additionally, the weight on the left-hand edge at every decision node is normalized to 1;
this invariant
ensures that WBDDs provide a canonical representation of Boolean functions.
We use the MQT DD package~\cite{Book:ZW2020} for backend WBDD support.
As distributed, MQT DD supports at most 128 qubits; we modified it to support up to $2^{31}$ qubits.

\paragraph{Symbolic Simulation.}
A symbolic simulation of a quantum circuit-computation \cite{sistla2022cflobdds,Book:ZW2020,9586191} uses a symbolic representation $qs$ of a quantum state and performs operations on $qs$ that correspond to quantum-circuit operations.
\begin{itemize}
    \item A quantum state of $n$ qubits is a vector of size $2^n \times 1$.
    Its entries are called \emph{amplitudes}, and the vector represents
    the probability distribution given by the squares of the absolute values of the amplitudes.
    In \sys, CFLOBDDs, BDDs, and WBDDs are used to represent functions of the form $f: \{ 0,1 \} ^n \rightarrow \mathbb{C}$---i.e., $f$ is a vector holding complex amplitudes.
    \item A quantum gate performs a linear transformation of a quantum state.
    Quantum-gate application is implemented by using a CFLOBDD, BDD, or WBDD to represent the matrix describing the quantum gate, and performing a matrix-vector multiplication (\cite[\S7.6--\S7.7]{sistla2022cflobdds},\cite{DBLP:journals/fmsd/BaharFGHMPS97}) of the gate matrix and the quantum state.
    \item For CFLOBDDs, BDDs, and WBDDs, operations like \texttt{prob}, \texttt{measurement\_counts},
    and \texttt{tail\_counts}
    are implemented as exact operations---i.e., no sampling---via projection and path-counting operations (\cite[\S7.8]{sistla2022cflobdds},\cite{toc:Bryant86}).
    For CFLOBDDs and BDDs, \sys computes \texttt{prob} via an efficient path-counting operation \cite[\S7.8.1 and \S10.1.2, respectively]{sistla2022cflobdds} to obtain the number of paths leading to each
    terminal value, and then projects the result onto the variables of interest (as specified by the user).
    \sys then returns the sum of the probabilities of the remaining paths.
    In the case of WBDDs as the backend, \sys computes the probability of every node (\cite[Ch.\ 5]{Book:ZW2020}) instead of
    counting paths.
    To compute \texttt{measurement\_counts}, \sys returns the number of paths that lead to the requested probability value
    within the provided threshold $\epsilon$.
    On querying \texttt{tail\_counts}, \sys returns the number of paths that lead to terminal values having probability $\texttt{prob} \geq \texttt{p}$, where \texttt{p} is the requested probability.
    \item 
    Once path-counts are computed,
    a measurement from the CFLOBDD, BDD, or WBDD symbolic representation of a quantum state is a data-structure traversal that can be carried out in time proportional to 
    $\mathcal{O}(\max(\textrm{number of qubits in the circuit}, \textrm{size of argument CFLOBDD}))$ 
\end{itemize}

\section{Experiments}
\label{Se:experiments}

In this section, we present some experimental results from using \sys on seven quantum benchmarks,
Greenberger–Horne–Zeilinger state creation (GHZ), Bernstein-Vazirani algorithm (BV), Deutsch-Jozsa algorithm (DJ), Simon's algorithm, Grover's algorithm, Shor's algorithm ($2n + 3$ qubits circuit by ~\cite{beauregard2002circuit}), and application of the Quantum Fourier Transform (QFT) to a basis state,
for different numbers of qubits.
Columns 2--4 of \tableref{quantum-table-detailed} show
the time taken for running the benchmarks with CFLOBDDs,
BDDs (CUDD 3.0.0 \cite{somenzi2012cudd}), and 
WBDDs (MQT DD
v2.1.0 \cite{zulehner2019package}).
For each benchmark and number of qubits, we created 50 random oracles and report the average time taken across the 50 runs.
For each run of each benchmark, we 
performed a measurement at the end of the circuit computation
and checked if the measured outcome is correct.
We ran all of the experiments on AWS machines: t2.xlarge machines with 4 vCPUs, 16GB memory, and a stack size of 8192KB, running on an Ubuntu OS.

One sees that CFLOBDDs scale better than BDDs and WBDDs for the GHZ, BV, and DJ benchmarks 
as the number of qubits increases.
BDDs perform better than CFLOBDDs and 
are comparable to WBDDs
for Simon's algorithm, whereas
WBDDs perform better than BDDs and CFLOBDDs
for QFT, Grover's algorithm, and Shor's algorithm.

We noticed that the BDD implementation suffers from precision issues;
i.e., if an algorithm with a large number of qubits contains too many Hadamard gates, it can lead to extremely low-probability values for each basis state, which are rounded to 0, which in turn causes leaves that really should hold different miniscule values to be coelesced unsoundly, leading to incorrect results.
To overcome this issue, one needs to increase the floating-point precision of the floating-point package used to represent BDD leaf values.
We increased the precision at 512 qubits ($^*$) and again at 2048 qubits ($^{**}$).

\begin{table}[!tb]
\centering
    \resizebox{.82\textwidth}{!}{
    \begin{tabular}{|c|c|c|c|c|c|c|c|}
    \hline
        \multirow{2}*{Benchmark} & \multirow{2}*{\#Qubits} &  
        CFLOBDD & BDD & WBDD & MQT DDSim & Quimb & GTN\\
        \cline{3-8}
        & & Time (sec) & Time (sec) & Time (sec) & Time (sec) & Time (sec) & Time (sec)\\
        \hline
        \multirow{10}*{GHZ} & 8 & 0.03 & 0.007 & 0.008 & 0.065 & 0.255 & 0.003\\
        \cline{2-8}
        & 16 & 0.03 & 0.008 & 0.011 & 0.068 & 0.368 & 0.010\\
        \cline{2-8}
        & 32 & 0.031 & 0.008 & 0.017 & 0.074 & 0.932 & \multirow{8}{*}{\minitab[c]{Memory \\ Error}}\\
        \cline{2-7}
        & 64 & 0.032 & 0.012 & 0.03 & 0.087 & 3.16 &\\
        \cline{2-7}
        & 128 & 0.035 & 0.026 & 0.06 & 0.116 & 12.1 &\\
        \cline{2-7}
        & 256 & 0.041 & 0.1 & 0.134 & \multirow{5}{*}{\minitab[c]{Not \\ Supported}} & \multirow{5}{*}{\minitab[c]{Memory \\ Error}} &\\
        \cline{2-5}
        & 512 & 0.053 & 0.552 & 0.35 & & &\\
        \cline{2-5}
        & 1024 & 0.078 & 3.01 & 1.05 & & &\\
        \cline{2-5}
        & 2048 & 0.13 & 18.8 & 3.59 & & &\\
        \cline{2-5}
        & 4096 & 0.239 & 129.92 & 13.33 & & &\\
        \hline
        \multirow{10}*{BV} & 8 & 0.037 & 0.007 & 0.007 & 0.068 & 0.288 & 0.005\\
        \cline{2-8}
        & 16 & 0.045 & 0.009 & 0.009 & 0.072 & 0.461 & 0.017\\
        \cline{2-8}
        & 32 & 0.06 & 0.013 & 0.012 & 0.082 & 1.21 & \multirow{8}{*}{\minitab[c]{Memory \\ Error}}\\
        \cline{2-7}
        & 64 & 0.095 & 0.033 & 0.019 & 0.105 & 4.64 &\\
        \cline{2-7}
        & 128 & 0.17 & 0.116 & 0.036 & \multirow{6}{*}{\minitab[c]{Not \\ Supported}} & 20.72 &\\
        \cline{2-5}\cline{7-7}
        & 256 & 0.33 & 0.42 & 0.082 & & \multirow{5}{*}{\minitab[c]{Memory \\ Error}} &\\
        \cline{2-5}
        & 512$^*$ & 0.68 & 2.12 & 0.235 & & &\\
        \cline{2-5}
        & 1024 & 1.43 & 10.65 & 0.753 & & &\\
        \cline{2-5}
        & 2048$^{**}$ & 3.1 & \multirow{2}{*}{\minitab[c]{Timeout \\ (15 min.)}} & 2.76 & & &\\
        \cline{2-3}\cline{5-5}
        & 4096 & 6.78 & & 10.77 & & &\\
        \hline
        \multirow{10}*{DJ} & 8 & 0.037 & 0.007 & 0.009 & 0.069 & 0.401 & 0.008\\
        \cline{2-8}
        & 16 & 0.045 & 0.01 & 0.012 & 0.075 & 0.873 & 0.034\\
        \cline{2-8}
        & 32 & 0.06 & 0.016 & 0.019 & 0.087 & 2.97 & \multirow{8}{*}{\minitab[c]{Memory \\ Error}}\\
        \cline{2-7}
        & 64 & 0.092 & 0.042 & 0.036 & 0.115 & 8.63 &\\
        \cline{2-7}
        & 128 & 0.16 & 0.17 & 0.082 & \multirow{6}{*}{\minitab[c]{Not \\ Supported}} & 43.53 &\\
        \cline{2-5}\cline{7-7}
        & 256 & 0.3 & 0.72 & 0.235 & & \multirow{5}{*}{\minitab[c]{Memory \\ Error}} &\\
        \cline{2-5}
        & 512$^*$ & 0.6 & 3.9 & 0.753 & & &\\
        \cline{2-5}
        & 1024 & 1.22 & 20.92 & 2.76 & & &\\
        \cline{2-5}
        & 2048$^{**}$ & 2.55 & \multirow{2}{*}{\minitab[c]{Timeout \\ (15 min.)}} & 10.77 & & &\\
        \cline{2-3}\cline{5-5}
        & 4096 & 5.55 & & 43.94 & & &\\
        \hline
        \multirow{5}*{Simons Alg.} & 4 & 0.05 & 0.014 & 0.008 & 0.064 & 0.272 & 0.004\\
        \cline{2-8}
        & 8 & 0.076 & 0.043 & 0.015 & 0.101 & 0.653 & 0.02\\
        \cline{2-8}
        & 16 & \multirow{3}{*}{\minitab[c]{Timeout \\ (15 min.)}} & 9.8 & 8.89 & 1.267 & 2.56 & \multirow{3}{*}{\minitab[c]{Memory \\ Error}}\\
        \cline{2-2}\cline{4-7}
        & 32 & & \multirow{2}{*}{\minitab[c]{Timeout \\ (15 min.)}} & \multirow{2}{*}{\minitab[c]{Timeout \\ (15 min.)}} & \multirow{2}{*}{\minitab[c]{Timeout \\ (15 min.)}} & 17.34 & \\
        \cline{2-2}\cline{7-7}
        & 64 & & & & & 267 &\\
        \hline
        \multirow{9}*{QFT} & 4 & 0.03 & 0.007 & 0.007 & 0.064 & 0.023 & 0.004\\
        \cline{2-8}
        & 8 & 0.04 & 0.043 & 0.009 & 0.068 & 0.035 & 0.012\\
        \cline{2-8}
        & 16 & 182.34 & 4.98 & 0.013 & 0.103 & 0.074 & 0.438\\
        \cline{2-8}
        & 32 & \multirow{6}{*}{\minitab[c]{Timeout \\ (15 min.)}} & \multirow{6}{*}{\minitab[c]{Timeout \\ (15 min.)}} & 0.027 & 0.154 & 0.231 & \multirow{6}{*}{\minitab[c]{Memory \\ Error}}\\
        \cline{2-2}\cline{5-7}
        & 64 & & & 0.104 & 0.363 & 1.64 &\\
        \cline{2-2}\cline{5-7}
        & 128 & & & 0.498 & \multirow{4}{*}{\minitab[c]{Not \\ Supported}} & 10.32 &\\
        \cline{2-2}\cline{5-5}\cline{7-7}
        & 256 & & & 2.73 & & 103.65 &\\
        \cline{2-2}\cline{5-5}\cline{7-7}
        & 512 & & & 17.54 &  & \multirow{2}{*}{\minitab[c]{Timeout \\ (15 min.)}} &\\
        \cline{2-2}\cline{5-5}
        & 1024 & & & 148.5 & & &\\
        \hline
        \multirow{4}*{Grovers Alg.} & 4 & 0.055 & 0.015 & 0.019 & 0.239 & \multirow{4}{*}{\minitab[c]{Memory \\ Error}} & \multirow{4}{*}{\minitab[c]{Memory \\ Error}} \\
        \cline{2-6}
        & 8 & 1.62 & 6.55 & 0.013 & 0.145 & & \\
        \cline{2-6}
        & 16 & \multirow{2}{*}{\minitab[c]{Timeout \\ (15 min.)}} & \multirow{2}{*}{\minitab[c]{Timeout \\ (15 min.)}} & 0.369 & 2.45 & &\\
        \cline{2-2}\cline{5-6}
        & 32 & & & \minitab[c]{Timeout \\ (15 min.)} & \minitab[c]{Timeout \\ (15 min.)} & &\\
        \hline  
        \begin{tabular}{@{}c@{}}
         Shor's Alg. \\
        (15, 2)
    \end{tabular} & 4 & \minitab[c]{Timeout \\ (15 min.)} & \minitab[c]{Timeout \\ (15 min.)} & 0.034 & 2.83 & \minitab[c]{Timeout \\ (15 min.)} & \minitab[c]{Timeout \\ (15 min.)}\\
    \hline
    \begin{tabular}{@{}c@{}}
         Shor's Alg. \\
        (21, 2)
    \end{tabular} & 5 & \minitab[c]{Timeout \\ (15 min.)} & \minitab[c]{Timeout \\ (15 min.)} & 0.252 & 9.35 & \minitab[c]{Timeout \\ (15 min.)} & \minitab[c]{Timeout \\ (15 min.)}\\
    \hline
    \begin{tabular}{@{}c@{}}
         Shor's Alg. \\
        (39, 2)
    \end{tabular} & 5 & \minitab[c]{Timeout \\ (15 min.)} & \minitab[c]{Timeout \\ (15 min.)} & 0.766 & 21.94 & \minitab[c]{Timeout \\ (15 min.)}  & \minitab[c]{Timeout \\ (15 min.)}\\
    \hline
    \begin{tabular}{@{}c@{}}
         Shor's Alg. \\
        (69, 4)
    \end{tabular} & 6 & \minitab[c]{Timeout \\ (15 min.)} & \minitab[c]{Timeout \\ (15 min.)} & \minitab[c]{Timeout \\ (15 min.)} & 204.08 & \minitab[c]{Timeout \\ (15 min.)} & \minitab[c]{Timeout \\ (15 min.)}\\
    \hline
    \begin{tabular}{@{}c@{}}
         Shor's Alg. \\
        (95, 8)
    \end{tabular} & 7 & \minitab[c]{Timeout \\ (15 min.)} & \minitab[c]{Timeout \\ (15 min.)} & \minitab[c]{Timeout \\ (15 min.)} & 192.05 & \minitab[c]{Timeout \\ (15 min.)} & \minitab[c]{Timeout \\ (15 min.)}\\
    \hline
    \begin{tabular}{@{}c@{}}
         Shor's Alg. \\
        (119, 2)
    \end{tabular} & 8 & \minitab[c]{Timeout \\ (15 min.)} & \minitab[c]{Timeout \\ (15 min.)} & \minitab[c]{Timeout \\ (15 min.)} & 206.62 & \minitab[c]{Timeout \\ (15 min.)} & \minitab[c]{Timeout \\ (15 min.)}\\
    \hline
 \end{tabular}
}
    \vspace{0.1in}
    \caption{ Performance of CFLOBDDs, BDDs, WBDDs using \sys; and other simulators like MQT DDSim, Quimb, and Google Tensor Network (GTN)
    }
    \label{Ta:quantum-table-detailed}
     \vspace{-0.4in}
\end{table}

Part of these results are similar to the work reported in~\cite{sistla2022cflobdds};
however, that paper did not use \sys.
The results of the present paper were obtained using \sys, and we also report results for WBDDs, as well as BDDs and CFLOBDDs (both of which were used in~\cite{sistla2022cflobdds}).
The numbers given in Table \ref{Ta:quantum-table-detailed} are slightly different from those given in \cite{sistla2022cflobdds}
because these quantum circuits exclusively use gate operations that are applied in sequence to the initial quantum state.
One can rewrite the quantum circuit to first compute
various gate-gate operations (either Kronecker product or matrix-multiplication operations) and then apply the resultant gate to the initial quantum state.
For example, consider a part of a circuit defined as follows:
\begin{center}
\begin{minipage}{0.6\textwidth}
\begin{minted}{python}
for i in range(0, n):
    qc.cx(i, n)
\end{minted}
\end{minipage}
\end{center}

Instead of applying CNOT ($cx$) sequentially for every $i$, one can construct a gate equivalent to $cx\_op = \Pi_{i=0}^{n-1}cx(i, n)$ and then apply $cx\_op$
to quantum state \texttt{qc} as follows:
\begin{center}
\begin{minipage}{0.6\textwidth}
\begin{minted}{python}
 cx_op = qc.create_cx(0, n)
 for i in range(1, n):
    tmp = qc.create_cx(i, n)
    cx_op = qc.gate_gate_apply(cx_op, tmp)
 qc.apply_gate(cx_op)
\end{minted}
\end{minipage}
\end{center}

\sys supports such operations as Kronecker product
and matrix product of two gate matrices.
\cite{sistla2022cflobdds} uses such computations for both oracle construction and as part of the quantum algorithm.~\tableref{quantum_benchmarks_2} shows the results on GHZ, BV, and DJ algorithms using the same circuit and oracle construction used in~\cite{sistla2022cflobdds}.
However, Simon's algorithm, Grover's algorithm, and Shor's algorithm in~\cite{sistla2022cflobdds} use operations outside \sys's computational model, and the results on these benchmarks differ from~\cite{sistla2022cflobdds}.
(Note that the results reported in \tableref{quantum_benchmarks_2} do not include the time taken for the construction of the oracle.)

\begin{table}[tb!]
    \centering
    \resizebox{0.55\textwidth}{!}{
    \begin{tabular}{|c|c|c|c|c|}
    \hline
        \multirow{2}*{Benchmark} & \multirow{2}*{\#Qubits} &  
        CFLOBDD & BDD & WBDD\\
        \cline{3-5}
        & & Time (sec) & Time (sec) & Time (sec)\\
        \hline
        \multirow{10}*{GHZ} & 8 & 0.03 & 0.008 & 0.009\\
        \cline{2-5}
        & 16 & 0.03 & 0.01 & 0.011\\
        \cline{2-5}
        & 32 & 0.034 & 0.035 & 0.017\\
        \cline{2-5}
        & 64 & 0.036 & 0.194 & 0.032\\
        \cline{2-5}
        & 128 & 0.04 & 1.47 & \multirow{6}{*}{\minitab[c]{Precision \\ Issue}}\\
        \cline{2-4}
        & 256 & 0.05 & 11.77 & \\
        \cline{2-4}
        & 512 & 0.07 & \multirow{4}{*}{\minitab[c]{Timeout \\ (15 min.)}} & \\
        \cline{2-3}
        & 1024 & 0.11 &  & \\
        \cline{2-3}
        & 2048 & 0.19 & & \\
        \cline{2-3}
        & 4096 & 0.36 &  & \\
        \hline
        \multirow{10}*{BV} & 8 & 0.001 & 0.001 & 0.001\\
        \cline{2-5}
        & 16 & 0.001 & 0.001 & 0.001\\
        \cline{2-5}
        & 32 & 0.002 & 0.006 & 0.001\\
        \cline{2-5}
        & 64 & 0.003 & 0.025 & 0.001\\
        \cline{2-5}
        & 128 & 0.005 & 0.089 & \multirow{6}{*}{\minitab[c]{Precision \\ Issue}}\\
        \cline{2-4}
        & 256 & 0.009 & 0.46 & \\
        \cline{2-4}
        & 512 & 0.015 & \multirow{4}{*}{\minitab[c]{Timeout \\ (15 min.)}} & \\
        \cline{2-3}
        & 1024 & 0.027 &  & \\
        \cline{2-3}
        & 2048 & 0.049 & & \\
        \cline{2-3}
        & 4096 & 0.086 & & \\
        \hline
        \multirow{10}*{DJ} & 8 & 0.005 & 0.001 & 0.001\\
        \cline{2-5}
        & 16 & 0.005 & 0.002 & 0.001\\
        \cline{2-5}
        & 32 & 0.005 & 0.006 & 0.001\\
        \cline{2-5}
        & 64 & 0.006 & 0.025 & 0.001\\
        \cline{2-5}
        & 128 & 0.006 & 0.084 & \multirow{6}{*}{\minitab[c]{Precision \\ Issue}}\\
        \cline{2-4}
        & 256 & 0.007 & 0.43 & \\
        \cline{2-4}
        & 512 & 0.008 & \multirow{4}{*}{\minitab[c]{Timeout \\ (15 min.)}} & \\
        \cline{2-3}
        & 1024 & 0.01 &  & \\
        \cline{2-3}
        & 2048 & 0.013 &  & \\
        \cline{2-3}
        & 4096 & 0.019 & & \\
        \hline
 \end{tabular}
}
    \vspace{0.1in}
    \caption{Performance of CFLOBDDs, BDDs, WBDDs using \sys on an alternate circuit implementation of GHZ, BV, DJ algorithms}
    \label{Ta:quantum_benchmarks_2}
\end{table}

We also compared \sys with three other quantum-simulation tools:
MQT DDSim~\cite{zulehner2019advanced}, Quimb~\cite{gray2018quimb}, and Google Tensor Network (GTN)~\cite{roberts2019tensornetwork}.
MQT DDSim is based on WBDDs (using MQT DD), whereas Quimb and GTN are based on tensor networks.
Their performance is shown in columns 6--8 of \tableref{quantum-table-detailed}.
Note that MQT DDSim does not support more than 128 qubits.
\section{Conclusion}
\label{Se:conclusion}

In this paper, we presented \sys, an extensible, open-source framework for quantum simulation using symbolic data-structures.
\sys supports CFLOBDDs and both unweighted and weighted BDDs as the underlying data-structures for representing quantum states and for performing quantum-circuit operations.
\sys is implemented as a Python library.
It provides an API to commonly used quantum gates and quantum operations, and also supports
operations for (i) computing the probability of a measurement leading to a given set of states, (ii) obtaining a representation of the set of states that would be observed with a given probability, and 
(iii) measuring an outcome from a quantum state.


\bibliographystyle{splncs04}
\bibliography{refs}

\appendix

\section{Installation Guide}
In this section, we provide steps for installing \sys.

\subsection{Prerequisites}
There are several prerequisites to install before installing and building \sys.

\begin{enumerate}
    \item Install make
    \item Download and install the Boost C++ library in the home directory
    \begin{lstlisting}[language=bash]
    wget https://boostorg.jfrog.io/artifactory/main/release/1.81.0/source/boost_1_81_0.tar.gz .
    tar -xvf boost_1_81_0.tar.gz
    \end{lstlisting}
    Export the BOOST\_PATH variable
    \begin{lstlisting}[language=bash]
    export BOOST_PATH="$HOME/boost_1_81_0"
    \end{lstlisting}
    \item Install mpfr
    \begin{lstlisting}[language=bash]
    sudo apt-get update
    sudo apt-get install libmpfr-dev
    \end{lstlisting}
    \item Install conda (if you want a virtual environment, recommended)
    \begin{lstlisting}[language=bash]
    wget https://repo.anaconda.com/miniconda/Miniconda3-latest-Linux-x86_64.sh .
    chmod a+x Miniconda3-latest-Linux-x86_64.sh
    bash Miniconda3-latest-Linux-x86_64.sh
    source ~/.bashrc
    conda create -n <env_name> python=3.9
    conda activate <env_name>
    \end{lstlisting}
    \item Set the Python include path to PYTHON\_INCLUDE variable
    \begin{lstlisting}[language=bash]
    export PYTHON_INCLUDE = <path_to_env>/include/python3.9
    \end{lstlisting}
    \item Install pip
    \item Install invoke
    \begin{lstlisting}[language=bash]
    pip install invoke
    \end{lstlisting}
    \item Install Pybind11
    \begin{lstlisting}[language=bash]
    pip install pybind11
    \end{lstlisting}
    \item Install gcc, g++
    \item Install autoconf
    \begin{lstlisting}[language=bash]
    sudo apt-get update
    sudo apt-get install autoconf
    \end{lstlisting}
\end{enumerate}

\subsection{Installing and Building \sys}
In this section, we list the steps to install and build \sys.
\begin{enumerate}
    \item Clone the repository and run a few init commands
    \begin{lstlisting}[language=bash]
    git clone https://github.com/trishullab/Quasimodo.git
    cd Quasimodo/
    git submodule update --init
    \end{lstlisting}
    \item Build the CUDD Library, which is used for BDDs
    \begin{lstlisting}[language=bash]
    cd cflobdd/cudd-complex-big/
    autoupdate
    autoreconf
    \end{lstlisting}
    \item Edit configure to add -fPIC flag to CFLAGS and CXXFLAGS and then, after saving the file, run the following commands:
    \begin{lstlisting}[language=bash]
    sed -i 's/: ${CFLAGS="-Wall -Wextra -g -O3"}/: ${CFLAGS="-Wall -Wextra -g -O3 -fPIC"}/g' configure
    sed -i 's/: ${CXXFLAGS="-Wall -Wextra -std=c++0x -g -O3"}/: ${CXXFLAGS="-Wall -Wextra -std=c++0x -g -O3 -fPIC"}/g' configure
    ./configure
    make
    cd ../..
    \end{lstlisting}
    \item Now build Quasimodo. (Edit the file tasks.py in python\_pkg to include -undefined dynamic\_lookup on line 25 if running on MACOS.)
    \begin{lstlisting}[language=bash]
    cd python_pkg/
    invoke build-quasimodo
    invoke build-pybind11
    \end{lstlisting}
\end{enumerate}

\noindent
\sys is now installed and can be used by importing it in any Python program.

\begin{itemize}
    \item We have provided 7 + 3 (using gate-gate operations) benchmarks that can be used as a starting guide on how to write programs using the \sys library.
    \item File quantum\_circuit.h provides an abstract class that shows all the supported list of gates and queries that are available in \sys.
\end{itemize}

\end{document}